\documentclass{article}
\usepackage{multirow}
\usepackage{INTERSPEECH2022}

\usepackage{setspace}
\title{THUEE SYSTEM DESCRIPTION FOR NIST 2020 SRE CTS CHALLENGE}

\name{
    Yu Zheng\textsuperscript{1}, Jinghan Peng\textsuperscript{1}, Miao Zhao\textsuperscript{1}, Yufeng Ma\textsuperscript{1}, Min Liu\textsuperscript{1} \\
    Xinyue Ma\textsuperscript{2}, Tianyu Liang\textsuperscript{2}, Tianlong Kong\textsuperscript{2}, Liang He\textsuperscript{2*}, Minqiang Xu\textsuperscript{1*}
}

\address{
\textsuperscript{1}SpeakIn Technologies Co. Ltd., ShangHai, China \\ \textsuperscript{2}Department of Electronic Engineering Tsinghua University, Beijing, China
}

\email{xuminqiang@speakin.ai, heliang@mail.tsinghua.edu.cn}

\begin{document}
\ninept
\maketitle
\begingroup\renewcommand\thefootnote{*}
\footnotetext{Corresponding author.}

\begin{abstract}
This paper presents the system description of the THUEE team for the NIST 2020 Speaker Recognition Evaluation (SRE) conversational telephone speech (CTS) challenge. The subsystems including ResNet74, ResNet152, and RepVGG-B2 are developed as speaker embedding extractors in this evaluation. We used combined AM-Softmax and AAM-Softmax based loss functions, namely CM-Softmax. We adopted a two-staged training strategy to further improve system performance. We fused all individual systems as our final submission. Our approach leads to excellent performance and ranks 1st in the challenge.
\end{abstract}



\section{Introduction}
\label{sec:intro}

The THUEE submission is the joint effort of the teams at SpeakIn Technologies Co. Ltd. and the department of electronic engineering Tsinghua university (THUEE). The subsystems, including ResNet74, ResNet152, and RepVGG-B2 were developed in this evaluation. All the subsystems contained a deep neural network followed by scoring and calibration. The ResNet74 and ResNet152 systems use cosine distance to score the trials. For each system, we describe the data involved in training, the system setup and the hyper-parameters. Finally, we report the experiment results of each subsystem and the fusion system on the SRE18 developing set, SRE19 evaluation set and SRE20 progress set.

\section{Datasets}
\label{sec:format}

\subsection{Training dataset}
\label{ssec:subhead}

The datasets used for training including:
\begin{itemize}
    \item Our own Chinese telephone data (CHI-tel).
    \item NIST SRE04-10.
    \item NIST SRE12 telephone data (SRE12-tel).
    \item NIST SRE16 telephone data (SRE16-tel).
    \item NIST SRE18 evaluation datasets.
    \item NIST SRE19 evaluation datasets.
    \item VoxCeleb 1+2.
    \item Switchboard phase1-3.
    \item MIXER6 telephone phone calls (MX6-tel).
    \item LibriSpeech.
\end{itemize}

In total, there are 255,214 speakers in the training datasets. We collected Chinese telephone data (CHI-tel) for a total of 230,000 speakers. We applied the following techniques to augment each utterance:
\begin{itemize}
    \item Reverberation: artificially reverberation using a convolution with simulated RIRs\cite{ko2017study} from the AIR dataset.
    \item Music: taking a music file (without vocals) randomly selected from MUSAN\cite{snyder2015musan}, trimmed or repeated as necessary to match duration, and added to the original signal (5-15dB SNR).
    \item Noise: MUSAN noises were added at one second intervals throughout the recording (0-15dB SNR).
    \item Babble: MUSAN speech was added to the original signal (13-20dB SNR).
\end{itemize}

We used all the data described above for training. All 16 kHz recordings were downsampled to 8 kHz using the FFmpeg tool. We used the 64-dimensional log Mel filter bank with energy based on Kaldi \cite{Povey_ASRU2011} with a 25 ms window size and a time shift of 10 ms. Mean normalization was applied using a moving window of 3 seconds. The Kaldi energy Voice Activation Detection (VAD) was used to detect speech activity.

\begin{table*}[t]
  \centering
  \caption{Results of the subsystems and fused system on the SRE18 dev cmn2 set, SRE20 progress and SRE20 test sets.}
  \label{table:resultspkin}
  \setlength{\tabcolsep}{3.1mm}{
\begin{tabular}{cccccccccc}
\hline
\multirow{2}{*}{System} & \multirow{2}{*}{Scoring} & \multicolumn{2}{c}{\textbf{SRE18 DEV CMN2}} & \multicolumn{3}{c}{\textbf{SRE20 PROGRESS SET}} & \multicolumn{3}{c}{\textbf{SRE20 TEST SET}} \\
 &  & EER & minDCF & EER & minDCF & actDCF & EER & minDCF & actDCF \\ \hline
ResNet74 & cos & \textbf{2.477} & \textbf{0.083} & 2.40 & 0.072 & 0.084 & - & - & - \\
ResNet152 & cos & 2.528 & 0.084 & 2.37 & 0.066 & 0.085 & - & - & - \\
RepVGG-B2 & cos & 2.560 & 0.103 & \textbf{2.12} & 0.068 & 0.082 & - & - & - \\ \hline
Fusion & cos & - & - & 2.23 & \textbf{0.063} & \textbf{0.076} & \textbf{2.53} & \textbf{0.061} & \textbf{0.068} \\ \hline

\end{tabular}
}
  \label{tab:a0}
  
\end{table*}

\subsection{Development dataset}
\label{ssec:subhead}
SRE19-dev dataset was used for fusion and calibration. SRE18-CMN2 was used for evaluation.
\section{SYSTEMS}
\label{sec:pagestyle}

\subsection{Backbone}
\subsubsection{ResNet}
\label{ssec:subhead}
As one of the most classical ConvNets, ResNet\cite{he2016deep} has proved its power in speaker verification. We used ResNet-74 and ResNet-152 in our systems. The block to build ResNet was bottleneck \cite{he2016deep}. The base channel was 64. The block numbers of ResNet74 and ResNet152 are (3, 4, 14, 3) and (3, 8, 36, 3) respectively.

\subsubsection{RepVGG}
In our previous work, we have proved that the RepVGG structure\cite{zhao2021speakin} is very effective in speaker recognition. We select RepVGG-B2 as our backbone. The model adopts 64 base channels. 

\subsection{Pooling Method}
The pooling layer aims to aggregate the variable sequence to an utterance-level embedding. We use the global statistics pooling (GSP) layer to aggregate the frame-level features along the time to obtain utterance-level representation. 

\subsection{Loss function}
\label{sssec:subsubhead}

Additive margin softmax (AM-Softmax) and angular softmax (AAM-Softmax) is proposed in\cite{wang2018additive} \cite{deng2019arcface}. In our system, we combine both of them, add penalty to both angle and angle cosine, and call it CM-Softmax. The loss function is defined as:
\begin{equation}
  L = -\frac{1}{n}\sum_{i=1}^n\log\frac{\mathrm{e}^{s[\mathrm{cos}(\theta_{y_i}+m_1)-m_2]}}{\mathrm{e}^{s[\mathrm{cos}(\theta_{y_i}+m_1)-m_2]} +\sum_{i\neq{}y_i}\mathrm{e}^{s\mathrm{cos\theta_j}}}
\end{equation}
where $\mathrm{cos}\theta_{y_i}=w_{y_i}^T f_i / \left \| w_{y_i} \right \|\left \| f_i \right \|$, $w_{y_i}$ is the weight vector of the class $f_i$ , and $f_i$ is the sample input. Also, $s$ is the scale factor and $m_1$ and $m_2$ are penalty margins for angle and angle cosine.

\subsection{Training Protocol}
\label{sssec:subsubhead}
Since most of our training data are Chinese speech data, there will be domain gaps during the evaluation phase. Therefore, the training process was divided into two stages. The first stage used all the training data to learn a pre-training model. The second stage performed model refinement \cite{Lamp88}, and only in-domain data was used to further fine-tune the pre-trained model from the first stage. The detailed parameters of training are as follows:

\begin{itemize}
\item \textbf{Stage 1: Pre-training}

All of the DNN architectures were trained using PyTorch \cite{paszke2019pytorch} with data parallelism over 8 Nvidia 3090 RTX GPUs. All data in section 2.1 were used for training. Therefore, the number of nodes in the last layer of the network was 255,214. We used SGD with momentum (set to 0.9) with a batch size of 320. Four seconds slice of audio was selected to train our model for each speech segment. 

In this stage, we adopted the ReduceLROnPlateau scheduler with a frequency of validating every 8,000 iterations, and the patience was set to 2. The minimum learning rate was 1.0e-6, and the decay factor is 0.5. Furthermore, the $m_1$ gradually increased from 0 to 0.2 \cite{Lamp91} and $m_2$ gradually increased from 0 to 0.1. Then a $model_{base}$ was selected. 

\item \textbf{Stage 2: Refinement}

The training data at this stage is a combination of NIST SRE04-10, NIST SRE18 eval, and NIST SRE19 eval, with a total of 4,596 speakers in total. And the number of nodes in the classifier is set to 4,596. Initialize the network with the $model_{base}$. Since the number of nodes in the last layer of the current network is different from that of  $model_{base}$,  and the training data is a subset of all datasets, extract the weight of the 4,596 speakers in $model_{base}$  as the initial weight of the current classifier. 

In this stage, we increased the chunk size to 1,000. $m_2$ is set to 0, and $m_1$ increases exponentially from 0.2 to 0.8 in 4,000 iterations. The detection frequency of the validation is 2,000 steps while the batch-size is set to 160.

\end{itemize}

\section{Scoring}
\label{sec:typestyle}

Originally, we used PLDA as the backend of the CNN network, and chose the SRE18 eval data and the SRE19 eval data as the backend training data. When we used a small-scale network, the results of the PLDA backend were better than the results of cosine. But with the increase of the model size and the addition of the refinement strategy, the result of PLDA degraded while the result of cosine score improved. Finally, the cosine result surpassed the one of PLDA. As a result, we only used cosine distance for scoring.

\section{Fusion and Calibration}
\label{sec:typestyle}

All of our individual systems were calibrated using logistic regression on the SRE19 dev data. The fusion was equal-weighted averages of the scores of all single systems.

\section{Results}
\label{sec:majhead}

Table \ref{table:resultspkin} presents the results of three subsystems trained for this challenge. Their performance is evaluated on the SRE18 dev CMN2 set and SRE20 progress set.  All of our subsystems were calibrated independently using logistic regression on the SRE19 dev data. The fusion system was an equal-weighted average of the scores of the systems. Finally, our system won first place in the NIST 2020 SRE CTS challenge.

The main problem identified in this challenge was the out-of-domain training speech corpora. The key to the problem was how to use the data when there were a few matching data. 


\bibliographystyle{IEEEbib}
\bibliography{refs,strings}

\end{document}